\documentclass[aps, prb, twocolumn, amsmath, amssymb, nofootinbib,superscriptaddress]{revtex4-2}
\usepackage{hyperref}
\hypersetup{
    colorlinks = true,
    linkcolor = [rgb]{0.70, 0.13, 0.13},
    citecolor = [rgb]{0.25, 0.41, 0.88},
    urlcolor = [rgb]{0.25, 0.41, 0.88}
}
\usepackage{amsmath}
\usepackage{amsfonts}
\usepackage{amssymb}
\usepackage{pifont}
\usepackage{mathtools}
\usepackage{bm}
\usepackage{cases}
\usepackage{braket}
\usepackage[version=4]{mhchem}
\usepackage{graphicx}
\usepackage{cleveref}
\allowdisplaybreaks%

\begin{document}

\title{Orbital homology of $p$ and $t_{2g}$ orbitals in models and materials}
 
\author{Gang v.~Chen}
\email{chenxray@pku.edu.cn}
\affiliation{International Center for Quantum Materials, School of Physics, 
Peking University, Beijing 100871, China}
\affiliation{Tsung-Dao Lee Institute, Shanghai Jiao Tong University, Shanghai 201210, China}
\affiliation{Collaborative Innovation Center of Quantum Matter, 100871, Beijing, China}
\affiliation{Beijing Key Laboratory of Quantum Devices, Peking University, Beijing 100871, China}

\author{Congjun Wu}
\email{wucongjun@westlake.edu.cn}
\affiliation{New Cornerstone Science Laboratory, Department of Physics, School of Science, 
Westlake University, Hangzhou 310024, Zhejiang, China}
\affiliation{Institute of Natural Sciences, Westlake Institute for Advanced Study, Hangzhou 310024, Zhejiang, China}
\affiliation{Institute for Theoretical Sciences, Westlake University, Hangzhou 310024, Zhejiang, China}
\affiliation{Key Laboratory for Quantum Materials of Zhejiang Province, School of Science, 
Westlake University, Hangzhou 310024, Zhejiang, China}

\begin{abstract}
The nominal divide between $p$- and $d$-electron systems often 
obscures a deep underlying unity in condensed matter physics. 
This review elucidates the {orbital homology} between 
the $p$ and $t_{2g}$ orbital manifolds, establishing the 
correspondence that extends from minimal model Hamiltonians to 
the complex behaviors of real quantum materials. We demonstrate 
that despite their distinct atomic origins, these orbitals host 
nearly identical hopping physics and spin-orbit coupling, formalized 
through an effective ${l=1}$ angular momentum algebra for the $t_{2g}$ case. 
This equivalence allows one to transpose physical intuition and 
theoretical models developed for $p$-orbital systems directly
onto the more complex $t_{2g}$ materials, and vice versa. 
We showcase how this paradigm provides a unified understanding of emergent phenomena, 
including non-trivial band topology, itinerant ferromagnetism, 
and unconventional superconductivity, across a wide range of platforms, 
from transition metal compounds, two-dimensional oxide heterostructures, 
and iron-based superconductors, to $p$-orbital 
ultracold gases. 
Ultimately, this $p$-$t_{2g}$ homology serves not only as 
a tool for interpretation but also as a robust design principle 
for engineering novel quantum states.
\end{abstract}
\maketitle

\section{Introduction}
\label{sec1}
 
In a fundamental sense, condensed matter physics and quantum materials 
are concerned with the microscopic constituents of physical degrees of freedom, 
their intrinsic properties, and the interactions among them. Understanding 
the microscopic nature of these physical degrees of freedom is therefore of 
paramount importance, and electron orbitals constitute one of the essential 
degrees of freedom for electrons~\cite{anderson1984basic,anderson1997concepts}.
Traditionally, the physics of $p$-electron systems (originating from ${l=1}$ orbitals),
and $d$-electron systems (${l=2}$) have been treated in largely separate domains. 
The former are central to the chemistry and physics of $sp$-bonded materials
like semiconductors, graphene, and molecular solids, 
while the latter dominate the rich and complex world of transition 
metal compounds~\cite{maekawa2004physics,khomskii2014transition}, 
renowned for phenomena such as high-temperature superconductivity, 
colossal magnetoresistance, and metal-insulator transitions. 
This conventional separation, however, obscures a profound and 
powerful underlying unity. An orbital homology exists 
between the $p$ orbital manifold and the $t_{2g}$ subset of $d$ orbitals, 
a correspondence that transcends their distinct atomic origins 
and provides a unified conceptual framework for understanding 
and designing quantum phenomena across a vast range of materials.

\begin{figure}[b]
\includegraphics[width=8cm]{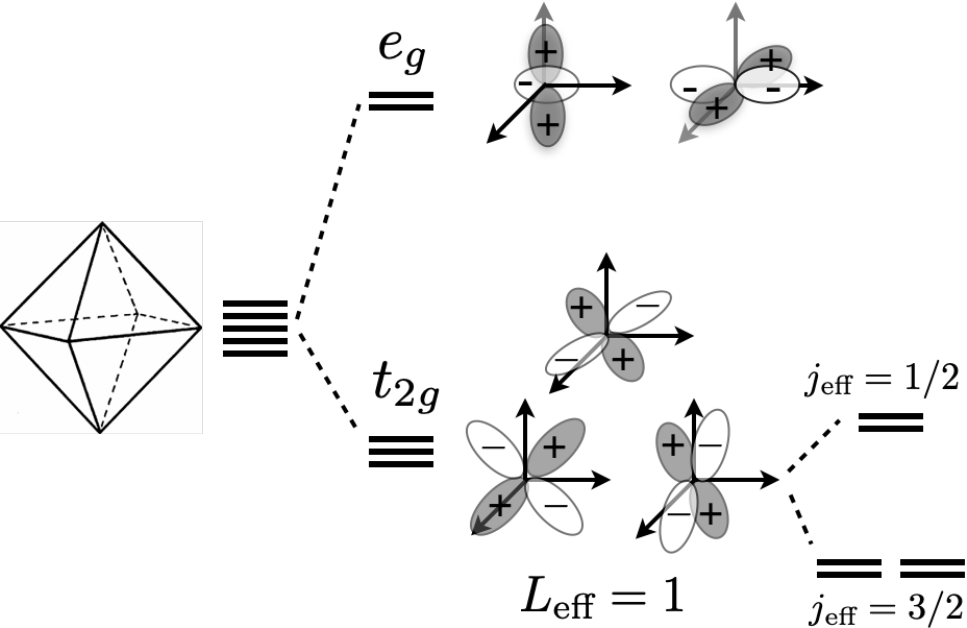}
    \caption{Schematic energy level diagram of $d$-electron splitting under an octahedral crystal field, followed by the effects of spin-orbit coupling. The crystal field lifts the degeneracy of the $d$-orbitals, splitting them into higher-energy $e_g$ and lower-energy $t_{2g}$ levels. With the inclusion of spin-orbit coupling, the $t_{2g}$ manifold further splits into a higher-energy ${j_{\text{eff}} = 1/2}$ doublet and a lower-energy ${j_{\text{eff}} = 3/2}$ quartet, corresponding to effective orbital angular momentum ${L_{\text{eff}} = 1}$. }
    \label{fig1}
\end{figure}

This equivalence is not merely a mathematical curiosity 
but a physical reality rooted in the fundamental principles of group theory. 
In an octahedral crystal field (see Fig.~\ref{fig1}), 
the five-fold degeneracy of the atomic $d$ orbitals is lifted, 
splitting them into a high-energy $e_g$ doublet ($d_{x^2-y^2}$, $d_{z^2}$) 
and a low-energy $t_{2g}$ triplet ($d_{xy}$, $d_{xz}$, $d_{yz}$). 
The transformational properties of these $t_{2g}$ orbitals under 
the symmetry operations of the octahedral group ($O_h$) are isomorphic 
to those of the $p$ orbital triplet ($p_x$, $p_y$, $p_z$). 
This algebraic homology permits a direct mapping:
\begin{equation}
d_{yz} \leftrightarrow -|p_x\rangle, \quad d_{xz} \leftrightarrow |p_y\rangle, 
\quad d_{xy} \leftrightarrow -|p_z\rangle.
\end{equation}
This mapping establishes that the $t_{2g}$ manifold behaves as 
an effective orbital angular momentum ${L_{\text{eff}} = 1}$ system 
(see Fig.~\ref{fig1}). 
The effective angular momentum formalism for the 
$t_{2g}$ manifold has a long history, 
dating back to the early crystal‐field theories 
of Van Vleck~\cite{VanVleck1932} 
and the paramagnetic resonance studies of Pryce~\cite{Pryce1950}, 
and was later systematized in the classic works of Griffith~\cite{Griffith1961}
and Abragam and Bleaney~\cite{AbragamBleaney1970}. 
In particular, the isomorphism between the 
$t_{2g}$ triplet and the 
p orbitals was recognized long ago and has been fruitfully exploited in 
the contexts such as magnetic resonance, Jahn-Teller physics, and optical spectroscopy. 
This foundational insight is the cornerstone of their homology, 
dictating parallel 
behaviours in two relevant regimes, {\sl i.e.} 
electronic hopping and spin-orbit coupling.

\begin{figure}
\includegraphics[width=6.5cm]{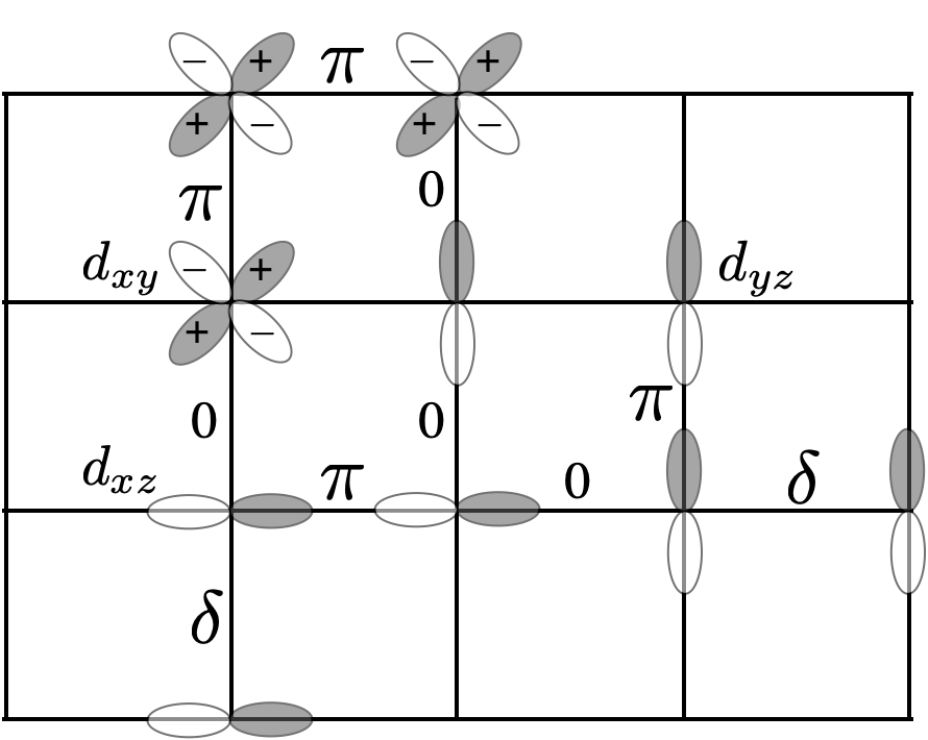}
\caption{Orbital-specific hoppings in a square lattice. $d_{xz}$ and $d_{yz}$ orbitals 
form $\pi$-bonds along $x$ and $y$ directions, respectively, while $d_{xy}$ exhibits 
the $\pi$ bonding along both $x$ and $y$ directions. The symmetry-demanded $0$ hopping
and the weak $\delta$ hopping are marked as well. }
\label{fig2}
\end{figure}

The orbital homology manifests most immediately in the kinetic energy of these electrons. 
The hopping amplitudes between orbitals on adjacent lattice sites are governed 
by the overlap of their wavefunctions, which is constrained by their symmetry and orientation. 
For instance, a $p_x$ orbital hops strongly along the $x$-direction through $\sigma$-bonding, 
but only weakly along $y$ via $\pi$-bonding. Similarly, a $d_{xz}$ orbital hops strongly 
along the $x$-direction with a strong $\pi$-overlap of its $xz$-lobe but negligibly 
along $y$ with a weak $\delta$-overlap in 2D (see Fig.~\ref{fig2}). 
This directional anisotropy is a hallmark of both manifolds.  
Consequently, the tight-binding models constructed for the $t_{2g}$ electrons 
in two-dimensional lattices
are quite analogous to those for the $p$ electrons in the same structure. 
This may allow researchers to leverage the more intuitive geometry of the $p$ 
orbitals to construct and solve models for the more complex $d$-electron systems, 
translating insights directly from one realm to the other.

The homology becomes much more useful with the inclusion of spin-orbit coupling (SOC), 
the key ingredient for many topological and/or correlated states. 
For atomic $p$ electrons, SOC takes the standard form 
$H_{\text{soc}} = \lambda \mathbf{L} \cdot \mathbf{S}$, 
where $\mathbf{L}$ is the genuine $l=1$ angular momentum operator. 
For the $t_{2g}$ electrons, the physical ${l=2}$ angular momentum 
is projected into the $t_{2g}$ subspace.  
Remarkably, the resulting operators $\mathbf{L}_{\text{eff}}$ 
obey the commutation relations of an ${l=1}$ algebra (see Fig.~\ref{fig1}). 
This leads to an identical SOC Hamiltonian for the $t_{2g}$ system, 
$H_{\text{soc}}^{t_{2g}} = - \lambda_{\text{soc}} \mathbf{L}_{\text{eff}} \cdot \mathbf{S}$, 
except the minus sign. 
This is a remarkable simplification. 
It implies that the complex SOC-driven physics in $t_{2g}$ materials, such as 
the formation of the celebrated $j_{\text{eff}} = 1/2$ ground state in strontium iridates Sr$_2$IrO$_4$~\cite{PhysRevB.57.R11039,PhysRevLett.102.017205,PhysRevB.78.094403,PhysRevLett.95.057205},
is directly analogous to the SOC physics in the $p$-orbital systems, though the former 
is more correlated.  
This shared platform enables the prediction and interpretation of topological insulators, 
Rashba effects, and topological superconductivity in $t_{2g}$ materials 
based on principles understood and explored in the $p$-orbital contexts.

The power of this paradigm is its ability to bridge the gap from the abstract models 
to real materials. We demonstrate this orbital homology across three principal domains.
In Sec.~\ref{sec2}, we explore the theoretical foundations and 
the material realizations of the orbital homology in topological and correlated phases. 
We begin with the Luttinger semimetal, showing how both $p$-orbital semiconductors 
and $t_{2g}$-based transition metal compounds exhibit identical quadratic band touching 
physics described by isomorphic Hamiltonians. 
We further mention how the Kane-Mele model and concepts of fragile topology 
find parallel realizations in both orbital systems, establishing a unified framework 
for understanding topological phase transitions across diverse material platforms 
including half-Heusler compounds and pyrochlore iridates.
Sec.~\ref{sec3} examines two-dimensional materials and interfaces, 
where the quantum confinement and symmetry breaking create ideal conditions 
for observing the functional equivalence. 
We analyze the oxide heterostructures such as LaAlO$_3$/SrTiO$_3$ and LaAlO$_3$/KTaO$_3$, 
showing how the emergent physics of the $t_{2g}$ orbitals mirrors that of the $p$-orbital 
systems in anisotropic hopping, Rashba spin-orbit coupling, and nematic superconductivity. 
The remarkable parallel between $t_{2g}$-based oxide interfaces and $p$-orbital topological 
insulators like Cu-doped Bi$_2$Se$_3$ highlights the universal role of spin-orbit coupling 
in driving unconventional superconducting states. 
The extension to iron-based superconductors is further elaborated. 
Finally, Sec.~\ref{sec4} investigates the ultracold atoms in optical lattices as 
pristine quantum simulators for $t_{2g}$ physics. 
We demonstrate how the $p$-orbital optical lattices can mimic the essential physics of
the $t_{2g}$ electron systems, enabling controlled studies of orbital ordering, magnetic frustration, 
and itinerant ferromagnetism without solid-state complications. 
This platform provides crucial insights into multi-orbital correlation effects 
and serves as an experimental testbed for theoretical predictions.
In Sec.~\ref{sec5}, we provide a summary about the current understanding 
and vision the future application of this $p$-$t_{2g}$ orbital correspondence.

\section{Orbital Homology in Topological and Correlated Phases}
\label{sec2}

The orbital homology between the $p$ and $t_{2g}$ orbital systems extends 
across the landscape of quantum materials, providing a more-or-less 
unified framework 
for understanding both topological phases and correlated electron phenomena. 
This section explores the homology between these orbital manifolds through 
their theoretical realizations in model Hamiltonians and their material manifestations 
in topological and correlated phases.

\subsection{Theoretical models}

The isomorphic orbital algebra shared by the $p$ and $t_{2g}$ manifolds means 
that minimal models for topological phases can be implemented on both $p$-orbital 
and $t_{2g}$-derived effective lattices with nearly identical outcomes. 
The Luttinger semimetal represents a fascinating class of quantum materials 
where the electron bands touch quadratically at discrete points in the Brillouin zone, 
creating unique electronic properties that bridge the gap between conventional metals 
and insulators. More significantly, the Luttinger semimetal is recognized 
as a parent state for numerous topological phases; through various perturbations, 
it can give rise to topological insulators, Weyl semimetals, Dirac semimetals, 
and other exotic states. In the modern nomenclature, the Luttinger semimetal is 
an example of topological crystalline semimetal~\cite{RevModPhys.93.025002,Gao_2019}. 
The physics of these Luttinger semimetallic systems finds remarkably parallel realizations 
in both $p$-orbital semiconductors and $t_{2g}$-based transition metal compounds, 
revealing the homology between these orbital manifolds.

\begin{figure}[t]
\includegraphics[width=6cm]{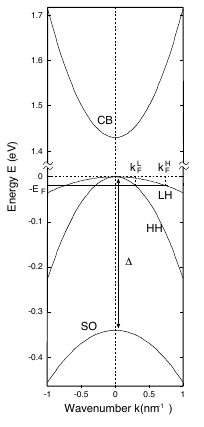}
\caption{Schematic picture of the band structure of GaAs near $k = 0$. 
CB means the conduction band, HH the heavy- hole band, LH the light-hole band, and SO the split-off band, respectively. When the small inversion symmetry breaking is neglected, all of them are doubly degenerate. 
The splitting $\Delta$ at $k = 0$ between the LH,HH and SO bands are due to the spin-orbit coupling. 
Figure is adapted from Ref.~\onlinecite{PhysRevB.69.235206}. }
\label{fig3}
\end{figure}

In the $p$-orbital semiconductor systems, particularly those with zincblende or diamond crystal structures, 
the Luttinger Hamiltonian emerges naturally from the coupling between the orbital and spin degrees of freedom. 
The derivation begins with the $p$-orbital basis ($p_x$, $p_y$, $p_z$) coupled with the electron spin, 
forming a six-dimensional Hilbert space. 
The strong spin-orbit coupling $H_{\text{soc}} = {\lambda} \mathbf{L} \cdot \mathbf{S}$ 
mixes these states, where $\mathbf{L}$ are the ${l=1}$ orbital angular momentum operators. 
When projected to the total angular momentum ${j=3/2}$ subspace, 
one obtains the celebrated Luttinger Hamiltonian~\cite{PhysRev.102.1030},
\begin{equation}
H_L = \frac{\hbar^2}{2m} \Big[ \gamma_1 k^2 - 2\gamma_2 \sum_{i} k_i^2 J_i^2 
- 4\gamma_3 \sum_{i<j} k_i k_j \{J_i, J_j \} \Big]
\end{equation}
where $\mathbf{J}$ represents the $4\times 4$ spin-3/2 matrices satisfying the SU(2) algebra, 
and the Luttinger parameters $\gamma_1$, $\gamma_2$, $\gamma_3$ 
characterize the material-specific band structure. Here, ${\{J_i,J_j \} =\frac{1}{2} (J_iJ_j + J_jJ_i) }$. 
The cubic anisotropy term with ${J_i,J_j}$ reflects the underlying crystal symmetry 
and leads to the characteristic band warping effects observed in materials 
like Ge, GaAs, $\alpha$-Sn, and HgTe~\cite{Bernevig_2006,PhysRevB.97.195139,PhysRevB.77.125319}. 
A schematic dispersion for Luttinger model is 
depicted in Fig.~\ref{fig3} for GaAs~\cite{PhysRevB.69.235206}.

Remarkably, an isomorphic mathematical structure emerges in the $t_{2g}$ electron systems 
despite their different orbital origin. Starting from the $t_{2g}$ manifold 
($d_{xy}$, $d_{xz}$, $d_{yz}$) with the strong spin-orbit coupling 
$H_{\text{soc}} = - \lambda_{\text{soc}} \mathbf{L}_{\text{eff}} \cdot \mathbf{S}$, 
the effective orbital angular momentum operators $\mathbf{L}_{\text{eff}}$  
obey the ${l=1}$ algebra due to the fundamental homology between the $t_{2g}$ and $p$ orbitals. 
Through the explicit mapping 
$d_{yz} \leftrightarrow -|p_x\rangle$, $d_{xz} \leftrightarrow |p_y\rangle$, 
$d_{xy} \leftrightarrow -|p_z\rangle$ in Sec.~\ref{sec1}, the $t_{2g}$ Hamiltonian transforms into,
\begin{equation}
H_{\text{eff}}^{t_{2g}} = \frac{\hbar^2}{2m_{\text{eff}}} \Big[ \tilde{\gamma}_1 k^2 
- 2 \tilde{\gamma}_2 \sum_{i} k_i^2 \tilde{J}_{i}^2 
- 4 \tilde{\gamma}_3 \sum_{i<j} \{ \tilde{J}_{i}, \tilde{J}_{j} \} k_i k_j \Big]
\end{equation}
where $\tilde{\mathbf{J}}$ represents the effective spin-3/2 operators 
in the $j_{\text{eff}}=3/2$ subspace, and the renormalized parameters $\tilde{\gamma}_i$ 
incorporate both the bare electron mass and correlation effects specific to $t_{2g}$ systems. 
 
\begin{figure}[t]
\includegraphics[width=8.5cm]{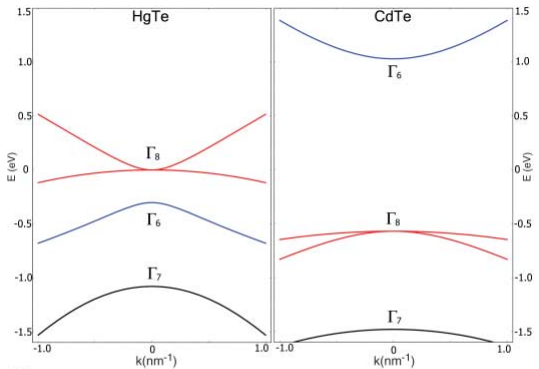}
\caption{Comparison of the bulk band structures of CdTe and HgTe at the $\Gamma$ point. In CdTe, the $\Gamma_6$ $s$-like conduction band lies above the $\Gamma_8$ $p$-like valence band. In HgTe, strong spin-orbit coupling inverts this order, placing the $\Gamma_8$ (heavy-hole) band above the $\Gamma_6$ band. The Luttinger model Hamiltonian, parameterized by $\gamma_1$, $\gamma_2$, and $\gamma_3$, captures the degenerate $\Gamma_8$ valence band structure in both semiconductors.
Figure is adapted from Ref.~\onlinecite{Bernevig_2006}. 
}
\label{fig4}
\end{figure}

While the Luttinger model itself only requires a symmetry-protected fourfold degeneracy at the 
$\Gamma$ point to describe its essential properties, 
this symmetry-based requirement overlooks the orbital character and microscopic 
origins of the underlying degrees of freedom. It is precisely the recognition that this model 
can be physically realized through both the genuine $j=3/2$ states of $p$-orbitals 
and the effective $j_{\text{eff}}=3/2$ 
states of the $t_{2g}$ orbitals that underscores the significance of their orbital homology.
Beyond the established realization of the Luttinger semimetal 
through the $j_{\text{eff}}=3/2$ states of $t_{2g}$ orbitals, 
it has been recently discovered that materials like pyrochlore iridate Pr$_2$Ir$_2$O$_7$ 
can host a Luttinger semimetal phase emerging from $j_{\text{eff}}=1/2$ states 
(see Fig.~\ref{fig4} and Sec.~\ref{sec2B})~\cite{Kondo_2015,PhysRevLett.111.206401,PhysRevX.8.041039}. In this scenario, 
the band index is partially provided by the pyrochlore's sublattice degree of freedom. 
This mechanism fundamentally transcends the conventional microscopic requirement of a 
$j=3/2$ angular momentum composition for the Luttinger model. 
Furthermore, leveraging the $p$-$t_{2g}$ equivalence, 
we can predict that $p$-orbital $j=1/2$ states on an appropriate non-Bravais lattice 
should similarly give rise to a Luttinger semimetal, 
generalizing the platform for this physics beyond its original conceptual boundaries.

The significance of this equivalence extends beyond the academic interest, as Luttinger semimetal serve as 
a versatile parent states for numerous topological phases~\cite{Kondo_2015,PhysRevLett.111.206401,PhysRevB.97.195139,PhysRevB.110.165120,PhysRevB.103.165139}. 
In both $p$-orbital and $t_{2g}$ realizations, breaking time-reversal symmetry 
through magnetic doping or external fields can transform the quadratic band touching into Weyl nodes, 
creating Weyl semimetals. Similarly, uniaxial strain or chemical pressure can lift the cubic symmetry, 
leading to Dirac semimetals or topological insulators depending on the perturbation direction and strength. 
The equivalence and homology ensure that the topological phase diagrams derived for $p$-orbital 
Luttinger semimetals directly inform the behavior of their $t_{2g}$ counterparts. 
Even for the non-topological models and phases, the orbital homology is expected to be quite useful. 
The spin-orbit-coupled metal has been proposed to understand the parent metallic 
state and the parity-breaking electronic nematic phase in the pyrochlore material Cd$_2$Re$_2$O$_7$ 
that is based on the spin-orbit coupling of the $t_{2g}$ 
electrons~\cite{PhysRevLett.115.026401,PhysRevLett.122.147602,Harter_2017}. 
The actual building-up of the spin-orbit-coupled metallic model with the Landau Fermi-liquid-like
approach, however, is quite general, and can be well suited for the $p$-orbital metals.

Although not as direct as the Luttinger semimetal, 
the Kane-Mele model~\cite{PhysRevLett.95.226801,PhysRevLett.95.146802}, 
a cornerstone of topological insulators, 
provides another compelling demonstration of this equivalence between $p$ and $t_{2g}$ orbitals. 
Originally conceived for spinful $p_z$ orbitals on a honeycomb lattice with spin-orbit coupling, 
it predicts the quantum spin Hall effect. Remarkably, the same topological phase can be realized in $t_{2g}$ systems. 
In materials such as the honeycomb iridate Na$_2$IrO$_3$~\cite{PhysRevLett.102.256403} 
or engineered oxide heterostructures~\cite{PhysRevLett.119.056803,PhysRevResearch.5.L022035,Meyer_2025,Ghiasi_2025}, 
the $t_{2g}$ orbitals on a honeycomb lattice, 
when subjected to strong spin-orbit coupling, 
were proposed to yield identical topological invariants and protected edge states, though Na$_2$IrO$_3$
was more popularly proposed as the Mott insulating Kitaev material due to the spin-orbit entangled $j=1/2$~\cite{PhysRevLett.105.027204}.
Similarly, concepts of fragile topology characterized by Wilson loop invariants 
find parallel realizations in both orbital families~\cite{PhysRevB.100.195135,PhysRevB.89.155114,PhysRevB.100.205126}. 
The nontrivial braiding of Wilson loops in the Brillouin zone, 
which signals fragile topology, depends on the specific orbital matrix elements and Berry connections. 
Since these are determined by the symmetry properties and algebra of the orbital basis, 
the identical transformation properties of $p$ and $t_{2g}$ orbitals guarantee 
that Wilson loop diagnostics yield equivalent results.

This underlying equivalence between the $p$-orbital and $t_{2g}$ realizations 
establishes a powerful paradigm for topological materials design, 
providing a robust conceptual framework that significantly accelerates 
the discovery and understanding of novel quantum states. The orbital homology and 
functional equivalence 
serve as a powerful design principle whereby theoretical predictions of topological 
phases in conceptually simpler $p$-orbital models can be directly translated 
to the more complex $t_{2g}$ materials, enabling efficient exploration 
of exotic quantum matter with tailored properties.

\subsection{Material realizations across orbital systems}
\label{sec2B}

The theoretical homology between the $p$ and $t_{2g}$ orbitals 
finds exquisite experimental realization across diverse material platforms, 
spanning from conventional semiconductors to complex correlated oxides. 
This material diversity illustrates the universal nature of orbital physics 
in determining the electronic behavior, demonstrating that the essential quantum phenomena 
are encoded not just in the specific atomic origins but also in the orbital characters.

\begin{figure}[t]
\includegraphics[width=6cm]{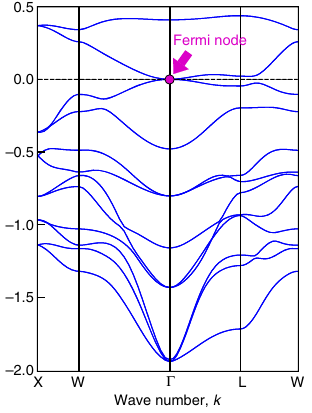}
\caption{The electronic band structure of Pr$_2$Ir$_2$O$_7$ from density functional theory (DFT). 
The Ir $t_{2g}$ electrons, under strong spin-orbit coupling, form $j_{\text{eff}}$ states. 
On the pyrochlore lattice, the four sublattices provide an internal degree of freedom that enables the formation of a quadratic band touching at the $\Gamma$ point, primarily derived from the $j_{\text{eff}}=1/2$ bands. This results in a Luttinger semimetal phase, where the low-energy physics is described by an effective Luttinger-type model but originates from the $j_{\text{eff}}=1/2$ manifold coupled to the lattice symmetry.
Energy unit is in eV. Figure is adapted from Ref.~\onlinecite{Kondo_2015}.
}
\label{fig5}
\end{figure}

In the $p$-orbital realm, half-Heusler compounds like LaBiPt and YPtBi provide 
exceptional platforms for studying the Luttinger semimetal physics 
with strong spin-orbit coupling~\cite{Nakajima_2015}. 
These materials exhibit the characteristic quadratic band touching predicted 
by the Luttinger Hamiltonian, serving as excellent testbeds for topological phases. 
Under appropriate perturbations (such as chemical substitution, hydrostatic pressure,
 or uniaxial strain), these $p$-orbital systems can be driven into various topological states 
 including Weyl semimetals and topological insulators. 
 Simultaneously, in the $t_{2g}$ domain, pyrochlore iridates such as Pr$_2$Ir$_2$O$_7$ 
and the related compounds display analogous quadratic band touching (see Fig.~\ref{fig5}) 
that can be similarly 
manipulated through external perturbations. The heavy $5d$ iridium ions in these materials 
provide exceptionally strong spin-orbit coupling, while the pyrochlore lattice geometry 
creates ideal conditions for realizing the effective $j_{\text{eff}}=1/2$ and $j_{\text{eff}}=3/2$
physics in Fig.~\ref{fig1}. Specifically, the Luttinger semimetal in Pr$_2$Ir$_2$O$_7$ 
arises mostly from the 
$j_{\text{eff}}=1/2$ manifold instead of the 
$j_{\text{eff}}=3/2$ manifold
of four pyrochlore sublattices. 
Experimental studies using angle-resolved photoemission spectroscopy have 
directly visualized the characteristic band degeneracies~\cite{Kondo_2015}, 
while transport measurements 
reveal the unique electronic responses expected for the Luttinger semimetals. 
The theoretical description of these iridate materials employs the same mathematical framework 
and topological invariants developed for $p$-electron topological insulators and semimetals, 
underscoring their hidden physical equivalence. The parallel presence of similar physics 
in both $p$-orbital and $t_{2g}$ material classes provides compelling evidence for their 
fundamental connection through orbital homology.

The comparative analysis across these material systems reveals universal physical signatures, 
including enhanced density of states near band touching points, 
non-Fermi liquid behaviors~\cite{PhysRevLett.111.206401},  
specific heat following distinctive power laws at low temperatures, 
and unusual magnetotransport behavior arising from non-trivial Berry curvature distributions.
Most importantly, both $p$-orbital and $t_{2g}$ systems exhibit similar responses 
to symmetry-breaking perturbations, following parallel pathways in their transitions 
to descendant topological and correlated phases. This universal behavior underscores 
that essential physics is dictated not by specific orbital origin but by symmetry properties 
and algebraic structure of the effective Hamiltonian, establishing the $p$-$t_{2g}$ 
correspondence as a powerful design principle for quantum materials exploration.

Beyond the Luttinger semimetal context, the $p$-$t_{2g}$ correspondence manifests powerfully 
in correlated electron systems. In heavy $5d$ transition metal oxides, 
notably the iridate family~\cite{Witczak_Krempa_2014}, 
the orbital homology illuminates the emergence of novel magnetic and topological states. 
The iridium $t_{2g}$ electrons experience competing energy scales: crystal field splitting 
($\sim$2-3 eV), Coulomb interactions ($\sim$1 eV), and remarkably strong spin-orbit coupling 
($\sim$0.4 eV). The spin-orbit coupling within the $t_{2g}$ manifold, 
mixes orbital and spin degrees of freedom to form total angular momentum $j_{\text{eff}}$ states 
(see Fig.~\ref{fig1}). 
The $t_{2g}^5$ electronic configuration of Ir$^{4+}$ ions 
gives rise to a $j_{\text{eff}} = 1/2$ ground state, conceptually analogous to 
the $j=1/2$ state that would arise from a $p^1$ configuration. 
This $j_{\text{eff}} = 1/2$ state exhibits unusual properties including anisotropic exchange interactions, canted antiferromagnetism, and proposed topological characteristics, 
and would in principle 
find natural relevance in the $p$-orbital analogue system 
if the strong correlation is present.

The power of the $p$-$t_{2g}$ homology extends to early transition metal oxides 
such as SrTiO$_3$~\cite{Gastiasoro_2020,PhysRevLett.12.474,PhysRevLett.125.087601,Nova_2019}. 
In SrTiO$_3$, the Ti $t_{2g}$ electrons display a rich phase diagram including 
superconductivity at low carrier densities, unusual ferroelectric-like behavior, 
and possible orbital ordering.
The $t_{2g}$ manifold's response to various perturbations, including strain, 
electric fields, and doping, closely mirrors how a $p$-orbital system would behave 
under equivalent symmetry-breaking conditions. The multi-band superconducting behavior 
and its sensitivity to disorder can be understood through the multi-orbital pairing symmetry 
derived from the $t_{2g}$ basis, 
which maps directly onto a $p$-orbital pairing problem. 
This perspective may be useful in explaining why certain scattering mechanisms 
strongly suppress superconductivity while others have minimal effect, 
based on the orbital character of the participating states. 


This unifying perspective reveals underlying connections between materials 
that would otherwise appear unrelated, and will stimulate cross-fertilization between research 
communities that traditionally focused on different material classes, 
accelerating progress in the understanding and design of complex quantum materials.
By recognizing that $t_{2g}$ systems can emulate the physics of $p$-orbital models, 
researchers can translate theoretical predictions from well-studied $p$-orbital contexts 
to the more complex realm of transition metal compounds,
except that the $t_{2g}$ systems can be more correlated than the $p$-orbital systems.
 The similar phenomenology observed in the $p$-electron systems like bismuth-based topological insulators 
and $t_{2g}$-based transition metal oxides, including their response to 
spin-orbit coupling, tendency toward nematic ordering, and unconventional 
superconducting instabilities, stems from their shared orbital properties. 
From the spin-orbit-driven magnetism of iridates~\cite{Caobook} to the complex orbital physics of
titanates~\cite{maekawa2004physics} and the topological states in various $4d$ and $5d$ compounds, 
this orbital homology illuminates the underlying mechanisms and emergent behaviors 
across diverse material families. 
By establishing this correspondence, we not only improve our understanding of the existing materials 
but also gain a useful framework for predicting and engineering new quantum states in transition metal compounds, harnessing the rich interplay between orbital, spin, charge, and lattice degrees of freedom that characterizes these materials.

\section{Two-Dimensional Materials and Interfaces}
\label{sec3}

 The functional equivalence and homology between the $p$ and $t_{2g}$ orbital systems manifest 
with particular clarity in low-dimensional structures and selected bulk materials, 
where quantum confinement, symmetry breaking, and specific crystal field environments 
create ideal conditions for observing this profound correspondence. 
Among these systems, the two-dimensional electron gases (2DEGs) formed at the interfaces of 
perovskite oxides, 
including the extensively studied LaAlO$_3$/SrTiO$_3$ system~\cite{Ohtomo2004,Hwang_Emergent_2012,Thiel2006,Cen2008,Bert2011,Reyren2007,Shunfeng2024} 
and 
the more recent LaAlO$_3$/KTaO$_3$ heterostructures~\cite{Liu2021,Chen2021,Baeumler2017,Li2023}, 
provide compelling platforms for exploring the deep connections 
between orbital physics and emergent quantum phenomena. 
Remarkably, this $p$-$t_{2g}$ equivalence extends beyond artificial heterostructures 
to bulk iron-based superconductors, where the tetrahedral crystal field and potential 
$e_g$-$t_{2g}$ level inversion in the FeX$_4$ (X = As, Se, Te) coordination environment 
enable topological band structures and even Majorana zero modes~\cite{Xu_2016,Zhang_2018,Wang_2018,Zhang_20182,Hao_2018}. 
These diverse quantum systems collectively reveal how the complex behavior of 
transition metal $d$-electrons can be understood through the more intuitive framework 
of $p$-orbital physics, while also highlighting remarkable parallels with topological 
insulator-based superconductors such as Cu-doped Bi$_2$Se$_3$~\cite{PhysRevLett.104.057001,PhysRevLett.107.217001,PhysRevLett.105.097001,Yonezawa_2016,PhysRevLett.108.107005,PhysRevLett.107.097001,PhysRevLett.108.057001,PhysRevB.90.100509}, 
establishing orbital homology as a useful principle across different material 
platforms and dimensionalities.


\subsection{Heterostructure interfaces}

\begin{figure}[t]
\includegraphics[width=8.5cm]{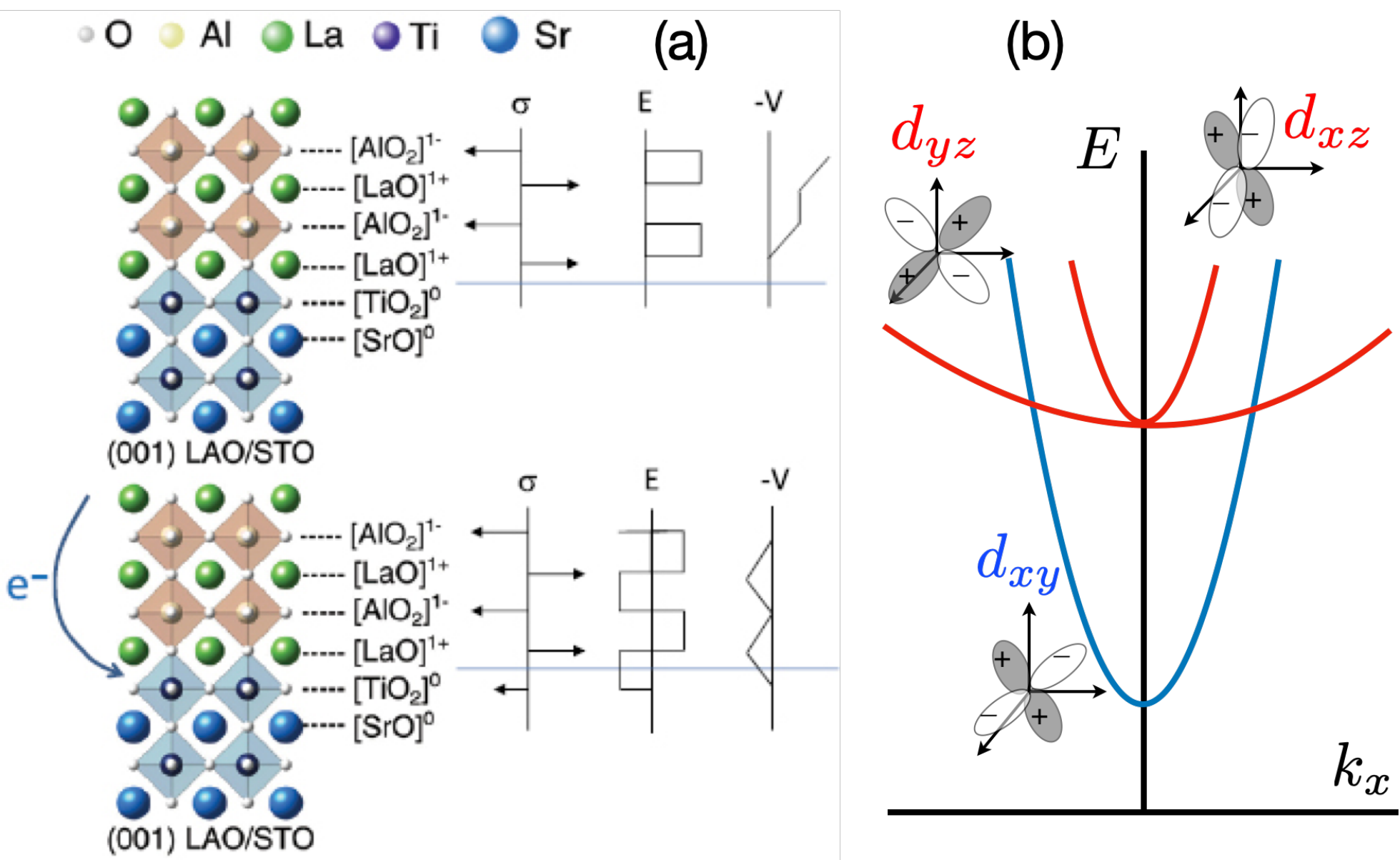}
\caption{
(a) Electronic reconstruction process at the (001) LaAlO$_3$/SrTiO$_3$ interface. The polar nature of LaAlO$_3$ results in a continuously growing electrostatic potential (polar catastrophe).To stabilize the system, a charge transfer of 0.5 electrons per unit cell occurs from the surface to the interface. This transfer screens the built-in electric field, leading to a stable potential profile and the formation of a conducting two-dimensional electron gas at the interface.
$\sigma$ is the planer electron density, $E$ is the surface electric field, and ${-V}$
is the electric potential. Figure is adapted from Ref.~\onlinecite{Rubano_2020}. 
(b) The dispersion of the $t_{2g}$ orbitals at the interface. }
\label{fig6}
\end{figure}

In the LaAlO$_3$/SrTiO$_3$ interface, the conduction electrons 
predominantly occupy the titanium $t_{2g}$ orbitals. 
The transition from bulk three-dimensional crystal to 
two-dimensional confinement induces a fundamental 
reorganization of the orbital energy landscape. 
While the $t_{2g}$ manifold remains degenerate in bulk SrTiO$_3$ 
under cubic crystal field symmetry, the interfacial environment 
breaks this degeneracy through two primary mechanisms,
the structural asymmetry normal to the interface plane 
and the electrostatic confinement potential that localizes electrons 
within a narrow quantum well. 
This confinement selectively affects the different $t_{2g}$ orbitals 
based on their spatial orientation relative to the interface. 
For the (001) interface, there exists a well-known interfacial electronic
reconstruction that creates a two-dimensional electron gas at the interface (see Fig.~\ref{fig6})~\cite{Rubano_2020,Ohtomo2004,Hwang_Emergent_2012}.
The $d_{xy}$ orbital, 
with its electron density concentrated within the interfacial plane, 
experiences the strongest quantum confinement effect.
 In contrast, the $d_{xz}$ and $d_{yz}$ orbitals, 
characterized by their out-of-plane spatial extension, 
respond less dramatically to the two-dimensional confinement.

More recently, the LaAlO$_3$/KTaO$_3$ heterostructure has 
emerged as a particularly intriguing system that exhibits 
even stronger spin-orbit coupling effects due to the heavy tantalum element. 
When fabricated along the (111) crystallographic direction, 
this interface displays remarkable superconducting properties 
below approximately 2K~\cite{Liu2021,Chen2021,Baeumler2017,Li2023}, 
with the superconducting state 
exhibiting clear signatures of nematic behavior~\cite{Zhang_2023}. 
The $t_{2g}$ orbitals in this geometry arrange in a triangular lattice 
that naturally hosts complex orbital textures,
while the strong atomic spin-orbit coupling of the 5$d$ tantalum 
electrons, significantly 
enhanced compared to the 3$d$ titanium electrons in SrTiO$_3$,
generates 
substantial mixing between the orbital and spin degrees of freedom. 
This strong spin-orbit coupling, 
operating within the $t_{2g}$ manifold with its effective ${l=1}$ character, 
plays a crucial role in stabilizing the unconventional superconducting state 
observed in these interfaces.

The emergent physics of the $t_{2g}$ orbitals in these oxide interfaces  
exhibits remarkable functional equivalence to the well-understood $p$-orbital systems. 
In particular, the $(xz,yz)$ doublet of the (001) interface behaves analogously 
to a two-dimensional $p_x$-$p_y$ system 
in its fundamental orbital symmetry and quasi-one-dimensional hopping characteristics, 
where the $d_{xz}$ orbital demonstrates strong hopping along the $x$-direction 
but significantly weaker hopping along the perpendicular $y$-direction,   
precisely mirroring the directional dependence of a $p_x$ orbital (see Fig.~\ref{fig2}). 
Simultaneously, the $d_{yz}$ orbital displays the complementary behavior, 
echoing the properties of a $p_y$ orbital. 
Furthermore, from the perspective of its response to structural inversion asymmetry 
and the resulting Rashba effect, the system behaves similarly to a $p_z$-like doublet
 under interfacial potential gradients.

The emergence of ferromagnetism at the LaAlO$_3$/SrTiO$_3$ interface~\cite{Bert2011} finds a fundamental explanation 
through the lens of $p$-$t_{2g}$ orbital homology~\cite{PhysRevLett.110.206401,PhysRevLett.112.217201}, revealing a universal mechanism 
for itinerant ferromagnetism in multi-orbital systems. Within the 2DEG, 
the $t_{2g}$-derived $(d_{xz}, d_{yz})$ doublet behaves as an effective $(p_x, p_y)$ orbital system. 
This orbital-selective hopping anisotropy creates distinct bandwidths and energy dispersions for different orbital states
(see Fig.~\ref{fig6}), a condition known to promote ferromagnetic instability in multi-orbital systems. 
Crucially, the presence of inter-orbital Hund's coupling provides the essential interaction that aligns electron spins across different orbital channels, while the anisotropic hopping ensures the kinetic energy gain through ferromagnetic alignment. The mechanism can be well described by the following multiple-orbital model~\cite{PhysRevLett.110.206401,PhysRevLett.112.217201},
\begin{eqnarray}
H &=& H_0+H_I , \label{eq4}\\
\mathcal{H}_0 &=& \sum_{\mathbf{k},\alpha} \frac{k_x^2 + k_y^2}{2m_0} a^{\dagger}_{0\alpha}(\mathbf{k}) a^{}_{0\alpha}(\mathbf{k}) 
\nonumber \\
&& \quad\quad\quad  
+ \sum_{\mathbf{k},\alpha,i=x,y} \left( \Delta + \frac{k_i^2}{2m_i} \right) a^{\dagger}_{i\alpha}(\mathbf{k}) a_{i\alpha}(\mathbf{k}) ,
\\
\mathcal{H}_I &=& U \sum_{\mathbf{r},i} n_{i\uparrow}(\mathbf{r}) n_{i\downarrow}(\mathbf{r})
+ U' \sum_{\mathbf{r},i \neq j} n_i(\mathbf{r}) n_j(\mathbf{r}),
\nonumber \\
&& \quad\quad\quad - J_H \sum_{\mathbf{r},i \neq j} \mathbf{S}_i(\mathbf{r}) \cdot \mathbf{S}_j(\mathbf{r}),
\end{eqnarray}
where \(a_{0\alpha}, a_{x\alpha}, a_{y\alpha}\) describe the \(xy\), \(xz\), and \(yz\) bands (with spin polarization \(\alpha = \uparrow, \downarrow\)), respectively.
$m_0$ is the effective mass for the $xy$ orbital, \(m_i\) ($i=x,y$) is the effective mass
for $yz$ and $xz$ orbitals, and \(\Delta\) is the subband crystal field splitting. 
We have assumed a tetragonal crystal field symmetry for the (001) interface in Eq.~\eqref{eq4}, 
so \(m_x = m_y\), 
and the one-dimensional hopping of the $xz/yz$ orbital is assumed as the limiting regime 
for the Luttinger liquid description. 
Moreover, \(n_{i\alpha}(\mathbf{r}) = a^{\dagger}_{i\alpha}(\mathbf{r}) a_{i\alpha}(\mathbf{r})\), 
\(n_{i}(\mathbf{r}) = \sum_{\alpha} n_{i\alpha}(\mathbf{r})\), and 
\(\mathbf{S}_{i}(\mathbf{r}) = \frac{1}{2} \sum_{\alpha\beta} a^{\dagger}_{i\alpha}(\mathbf{r}) \bm{\sigma}_{\alpha\beta} a_{i\beta}(\mathbf{r})\). 
As usual, one expects the intra-orbital interaction Hubbard \(U\) to be the largest interaction, 
with the inter-orbital interaction \(U^{\prime}\) and the Hund's coupling \(J_{H}\) smaller.

Within these quasi-1D bands, the electron-electron interactions are not adequately described by Fermi liquid theory. Instead, the low-energy physics is governed by the Luttinger liquid framework, where spin and charge degrees of freedom separate, and the spin fluctuations are significantly enhanced. The crucial ingredient is the Hund's coupling, $J_H$, which operates on the Ti sites. In this multi-orbital environment, Hund's rule favors the alignment of electron spins across the different quasi-1D channels (e.g., in the $d_{xz}$ and $d_{yz}$ orbitals)
and the $d_{xy}$ orbital. 
The enhanced spin susceptibility inherent to the Luttinger liquid state in these 1D bands amplifies this effect, 
providing a highly favorable environment for the ferromagnetic alignment of spins to lower the total energy via the Hund's interaction. This synergy between multi-orbital Hund's coupling and the unique correlation physics of quasi-1D Luttinger liquids offers a robust and distinct pathway to stabilize the interfacial ferromagnetic state observed in 
the LaAlO$_3$/SrTiO$_3$ interface. Even slightly away from the Luttinger liquid limit, 
the argument for the ferromagnetism is expected to be valid. 
This mechanism finds rigorous mathematical foundation in the exact solutions of
the isomorphic $p$-orbital models on similar lattice geometries (see Sec.~\ref{sec4})~\cite{PhysRevLett.112.217201}. 
This combined effect of orbital anisotropy and Hund's coupling, operating within the framework of 
$p$-$t_{2g}$ homology, thus establishes a universal paradigm for understanding ferromagnetic states across 
diverse material platforms, from oxide interfaces to engineered $p$-orbital systems.

Another striking parallel emerges when comparing the superconducting properties 
of these $t_{2g}$-based oxide interfaces with those of Cu-doped Bi$_2$Se$_3$, 
a $p$-orbital topological insulator that becomes superconducting 
upon copper intercalation~\cite{PhysRevB.97.041408,Zhang_2023,Yonezawa_2016,Hecker_2018,Matano_2016}. 
Both systems exhibit nematic superconductivity, 
characterized by the spontaneous breaking of rotational symmetry in the superconducting state. 
In Cu-doped Bi$_2$Se$_3$, the strong spin-orbit coupling inherent to the $p$-orbital-derived 
bands of the bismuth selenide matrix stabilizes an odd-parity pairing state that transforms  
non-trivially under crystal rotations. Similarly, in the LaAlO$_3$/KTaO$_3$ (111) interface, 
the $t_{2g}$ electrons, subjected to strong spin-orbit coupling and confined 
in a triangular lattice geometry, develop a superconducting order parameter 
that breaks the threefold rotational symmetry of the underlying crystal structure. 
This remarkable similarity between systems based on fundamentally different orbital 
types, $p$ orbitals in Cu-doped Bi$_2$Se$_3$ and $t_{2g}$ orbitals in KTaO$_3$ 
interfaces, highlights the universal role of spin-orbit coupling 
in driving nematic superconductivity, regardless of the specific orbital 
origin of the conducting states.  
This connection is especially clear once the superconducting properties 
are understood from the perspective of the more universal and generic
Ginzburg-Landau theory.

The broader implications of this understanding extend to a growing family of 
complex oxide interfaces and topological materials. 
The $p$-$t_{2g}$ correspondence provides a powerful conceptual framework for 
interpreting experimental observations and guiding materials design across 
diverse quantum systems. This perspective has proven particularly valuable 
in explaining the evolution of electronic properties with layer thickness, 
gate voltage, and strain—parameters that directly modulate the quantum 
confinement and orbital polarization in these systems. 
By mapping the complex $t_{2g}$ orbital physics onto the more intuitive language of $p$-orbital systems, researchers can leverage well-established theoretical tools and physical intuitions to engineer quantum phenomena in these technologically promising materials, from topological phases to unconventional superconductivity. 
The deep connections between seemingly disparate materials like oxide heterostructures 
and topological insulators, as exemplified by the parallel nematic superconductivity in LaAlO$_3$/KTaO$_3$ interfaces and Cu-doped Bi$_2$Se$_3$, underscore the fundamental nature of orbital-structure relationships 
in determining emergent quantum behavior, opening new pathways for oxide electronics and quantum information applications.

\subsection{Iron-based superconductors}

\begin{figure}[b]
\includegraphics[width=8.5cm]{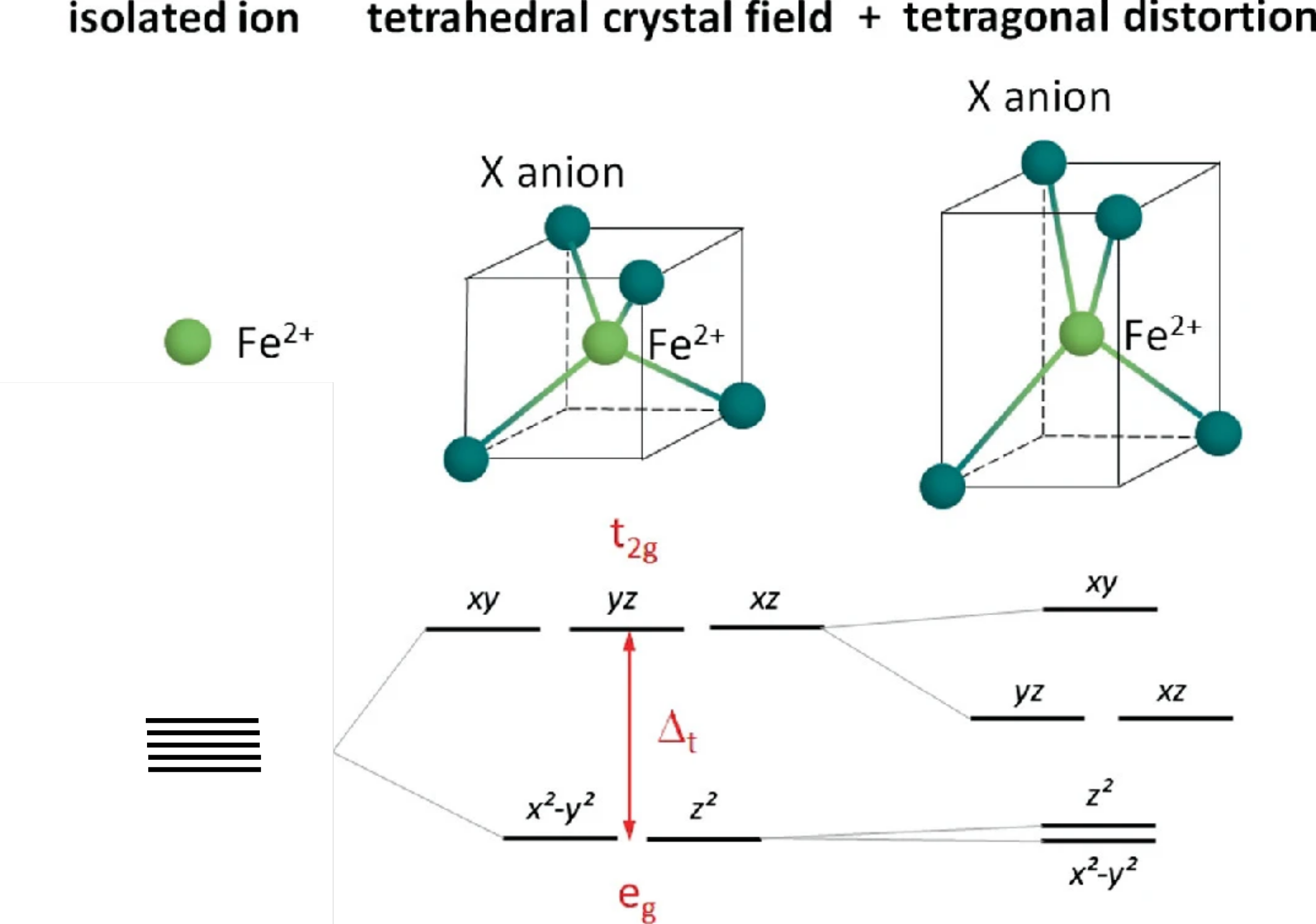}
\caption{Crystal field levels of the Fe $3d$ orbitals
in Fe-based superconductors due to the tetrahedral crystal field 
and an additional tetragonal distortion. Figure is adapted from
Ref.~\onlinecite{Haindl2021} with modification.}
\label{fig7}
\end{figure}

The exploration of topological phenomena in low-dimensional systems, 
particularly at interfaces and in two-dimensional materials as discussed 
in previous subsections, finds a remarkable parallel in the bulk 
and two-dimensional electronic structure of iron-based superconductors. 
This connection is again rooted in the fundamental orbital homology 
between the iron $t_{2g}$ orbitals and $p$ orbitals. In iron-based superconductors~\cite{Zhang_2018}, 
the iron atoms reside in a tetrahedral coordination environment     
formed by the pnictogen or chalcogen atoms, creating a crystal field 
that splits the $3d$ orbitals into $e_g$ and $t_{2g}$ manifolds
that are reversed from the octahedral environment. 
Despite this little difference, the $t_{2g}$ orbitals still behave
as active orbitals with active spin-orbit couplings (see Fig.~\ref{fig7}).
The orbital homology between $p$ and $t_{2g}$ orbitals 
manifests through two deeply connected physical phenomena, the microscopic mechanism of spin-orbit coupling driven band inversion and the subsequent emergence of topological superconductivity with Majorana zero modes. 
This dual demonstration reveals the $p$-$t_{2g}$ equivalence as a fundamental 
organizing principle governing topological phenomena across disparate material classes.

The theoretical foundation of this homology lies in the similar transformation properties 
of the $t_{2g}$ orbital triplet under the glide-mirror and inversion symmetries of 
the P4/nmm space group
as the behaviors of the $p$ orbital electrons in the relevant materials~\cite{Hao_2018}. 
This algebraic equivalence establishes the $t_{2g}$ manifold as an effective orbital 
angular momentum ${l_{\text{eff}} = 1}$ system, directly analogous to the intrinsic ${l = 1}$ 
character of the $p$ electrons. This mapping becomes particularly significant in materials 
like Fe(Te,Se) and Li(Fe,Co)As, where the combination of tetrahedral coordination, 
optimized anion height, and strong electron correlations creates conditions favorable 
for topological band inversions within the $t_{2g}$-derived electronic states~\cite{Hao_2018}.

The physical manifestations of this orbital homology are profound. 
The low-energy effective models describing these $t_{2g}$-based band inversions,
whether in monolayer FeSe/SrTiO$_3$ or bulk Fe(Te,Se), are formally identical to the 
Bernevig-Hughes-Zhang model originally developed for $p$-type HgTe/CdTe quantum wells. 
There, strong spin-orbit coupling induces a fundamental band inversion 
between the $s$-type ($\Gamma_6$) and $p$-type ($\Gamma_8$) bands (see Fig.~\ref{fig4}), 
with the topological nature captured by the celebrated Bernevig-Hughes-Zhang (BHZ) model~\cite{Bernevig_2006},
where the mass term changes sign at the critical thickness, 
signaling the topological phase transition tuned by the spin-orbit coupling.  
Remarkably, this identical physical picture emerges in the iron-based superconductors, 
albeit through the agency of $t_{2g}$ orbitals acting as the effective $p$ orbitals. 
This
equivalence enables the construction of nearly identical low-energy $\mathbf{k} \cdot \mathbf{p}$ models around high-symmetry points. For instance, at the $\Gamma$ point, the effective Hamiltonian takes the form~\cite{Bernevig_2006,Hao_2018},
\begin{equation}
H_\Gamma(\mathbf{k}) = \epsilon_0(\mathbf{k}) + \begin{bmatrix}
-M(\mathbf{k}) & A k_+ & 0 & 0 \\
A k_- & M(\mathbf{k})  & 0 & 0 \\
0 & 0 & -M(\mathbf{k}) & - Ak_- \\
0 & 0 & -Ak_+ & M(\mathbf{k}) 
\end{bmatrix} ,
\end{equation}
where $M(\mathbf{k}) = M - B(k_x^2 + k_y^2), k_{\pm} = k_x \pm i k_y$, precisely mirroring the BHZ form. 
The band inversion condition $MB > 0$, now occurring between $t_{2g}$-derived states of opposite parity, 
demonstrates conclusively that the topological band inversion mechanism transfers completely across 
two orbital families, and the topological transition is again driven by the 
spin-orbit coupling that is operates equivalently on the $t_{2g}$ orbitals as 
the $p$ orbitals. 
This theoretical equivalence translates directly to the parallel phenomena.
The Dirac-cone surface states and helical spin textures observed in Fe(Te,Se) via 
ARPES represent 
the topological physics~\cite{Zhang_2018}, that was first observed in the $p$-orbital systems, 
now emerging from the effective $p$-wave character of the $t_{2g}$ electrons.

This orbital homology extends powerfully into the realm of topological superconductivity and Majorana physics. 
In conventional semiconductor-based approaches, Majorana bound states are engineered through proximity effects 
between $s$-wave superconductors and topological insulators or semiconductors 
with strong Rashba-type $p$-orbital character~\cite{PhysRevLett.100.096407,PhysRevLett.105.077001}.
Iron-based superconductors, however, achieve this topology intrinsically 
through a remarkable ``self-proximity effect" that leverages their multi-orbital nature~\cite{Hao_2018}. 
The topological surface states emerging from the $t_{2g}$-band inversion, 
themselves possessing the $p$-orbital character, become naturally proximitized 
by the intrinsic $s_\pm$-wave bulk superconductivity of the same material. 
This creates an inherent platform for topological superconductivity 
described by an effective $p$-wave pairing potential, without requiring artificial heterostructures.

The experimental observations of zero-bias conductance peaks localized in the vortex cores of FeTe$_{0.55}$Se$_{0.45}$, exhibiting key characteristics of Majorana bound states including non-splitting behavior with spatial position and temperature scaling consistent with topological protection, provides compelling evidence that the $t_{2g}$ electrons can emulate the complete topological pathway from band inversion to Majorana physics~\cite{Wang_2018,Zhang_20182,Zhang_2018}. This dual demonstration, encompassing both the microscopic SOC-driven band inversion mechanism and the subsequent emergence of topological superconductivity with non-Abelian excitations, establishes the $p$-$t_{2g}$ orbital homology not as a mere mathematical curiosity but as a robust physical principle capable of guiding the discovery and design of correlated topological matter across the quantum materials landscape.

\section{Ultracold Atoms in Optical Lattices as Quantum Simulators}
\label{sec4}

The study of orbital physics in condensed matter systems faces numerous challenges, 
including disorder, complex interactions, and the difficulty of precisely controlling
material parameters. Ultracold atoms in optical lattices provide an exceptionally clean 
and highly tunable alternative platform for exploring the fundamental
physics of orbital systems, particularly the functional equivalence 
and the homology between $p$- and $t_{2g}$-orbital manifolds. 
These quantum atomic matter states offer unprecedented control over lattice geometry, 
interaction strength, and orbital degrees of freedom, 
enabling the simulation of complex orbital phenomena 
that are difficult to isolate in solid-state systems.

Optical lattices created by interfering laser beams
can be engineered to produce various lattice geometries,
including the cubic, hexagonal, and triangular structures.
For the fermionic atoms in $p$-orbitals, 
they directly simulate $p$-orbital physics with remarkable fidelity. 
The anisotropic nature of $p$-orbital wavefunctions,
with their directional lobes and orientation-dependent hopping amplitudes,
naturally emerges in these systems. 
More significantly, the Hamiltonian 
in a cubic or square optical lattice 
can be precisely~\cite{PhysRevLett.112.217201} 
tuned to mimic the essential physics 
of $t_{2g}$ electron systems in transition metal oxides. 
The $p_x,p_y$-orbital bands in the honeycomb lattice exhibit both flat 
and topological band structures~\cite{Wu_2007},
which can also be mapped into both the $t_{2g}$ and even $e_g$-orbitals 
in transition metal oxides~\cite{xusl2022}.
This mapping is possible because both systems share 
the same fundamental orbital symmetry and transformation properties 
under the point group operations.

The key advantage of optical lattice simulations lies in their ability 
to isolate specific aspects of the $t_{2g}$ physics while avoiding complications 
such as disorder, phonon scattering, and complex interactions that often obscure 
the fundamental physics in material systems. 
Researchers can systematically investigate phenomena such as orbital ordering, 
geometric frustration, and spin-orbit coupling effects by controlling 
experimental parameters including lattice depth, interaction strength 
via Feshbach resonances, and artificial gauge fields that simulate 
spin-orbit coupling~\cite{Han2019Physics}. 
This clean tunability makes optical lattices an 
ideal testbed for theoretical predictions about
$t_{2g}$ systems before attempting to observe them in solid-state materials. 

\subsection{Homology in flat bands}

\begin{figure}
\includegraphics[width=8.5cm]{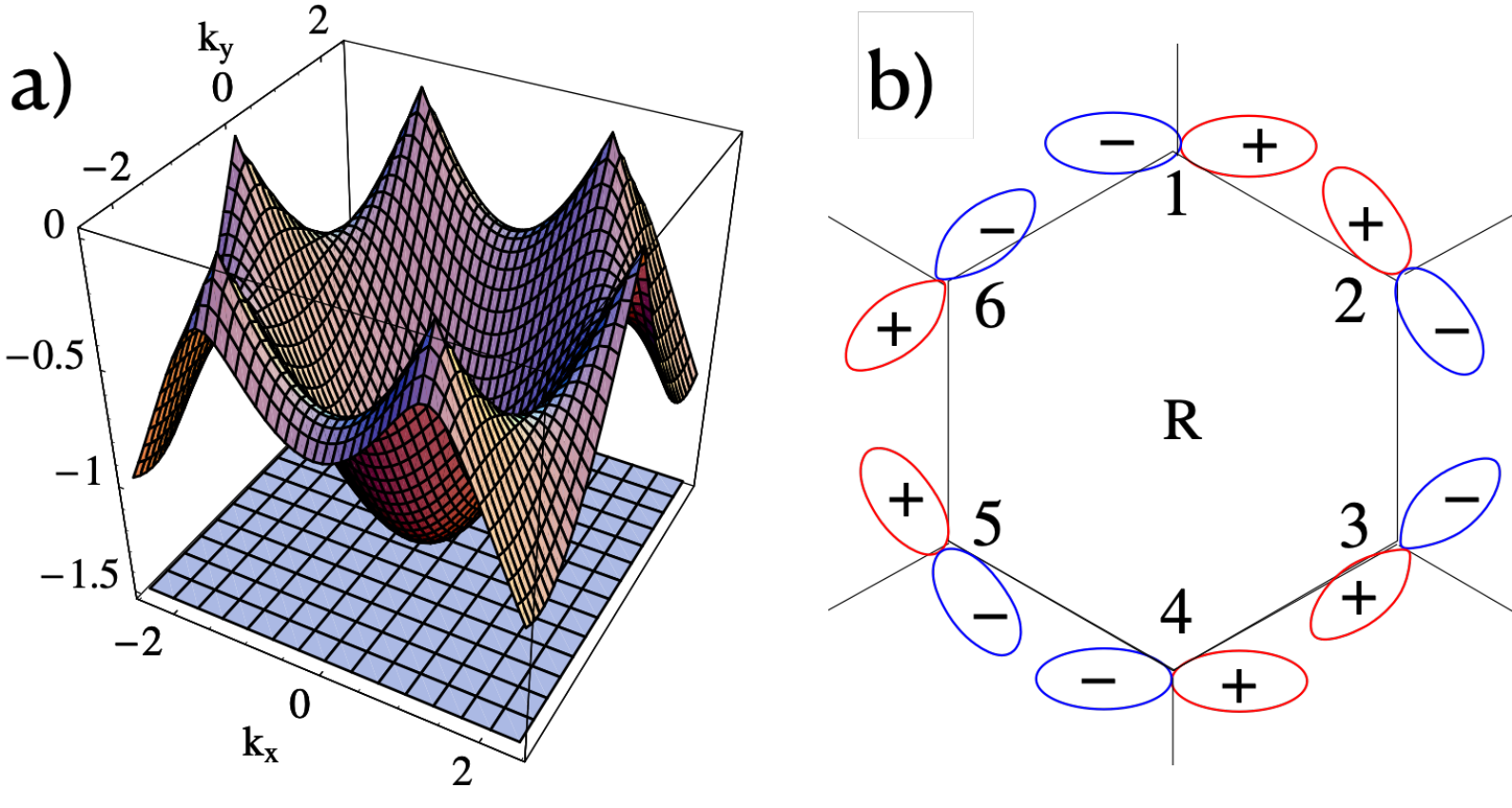}
\caption{a) The energy spectra of the $p_x$ and $p_y$-orbital bands,
including the bottom flat-band and dispersive Dirac band. The spectra
are symmetric with respect to zero, and only the lower two bands are shown.
b) The localized eigenstate for the flat band.
Adapted from Ref.~\onlinecite{Wu_2007}.}
\label{fig:flatdirac}
\end{figure}

As shown in Fig.~\ref{fig2}, one aspect of the $p$-$t_{2g}$ orbital homology 
lies in their hopping integrals that are directly responsible 
for their multi-orbital band structures (including the emergence of flat bands) 
on different lattices.
In particular, for the $p_x$-$p_y$-orbital bands 
in the honeycomb lattice, they are markedly different from that of graphene 
as shown in Fig.~\ref{fig:flatdirac} (a).
There are four bands with the lower two shown in Fig.~\ref{fig:flatdirac}(a), 
and the dispersion of the other two are opposite to the lower ones due to the particle-hole symmetry. 
The bottom and top bands are flat.
They can be constructed as superpositions of a set of degenerate localized eigenstates 
depicted in Fig.~\ref{fig:flatdirac} (b), which exists in each hexagonal plaquette 
as a combined effect of destructive quantum interference and hopping anisotropy. 
The exquisite control in a clean, tunable environment allows for the isolated study 
of flat-band-driven correlation physics, such as Wigner crystallization~\cite{wu2008a} 
or itinerant ferromagnetism~\cite{zhangSZ2010}.
The middle two bands exhibit the same dispersion as in graphene also with Dirac cones, 
but their wavefunctions are very different.
They provide a new mechanism for large topological gap opening beyond 
the $s$-$p$ band inversion mechanism~\cite{zhanggf2014}. 
The flat and dispersive bands exhibit quadratic touching, 
which can be lifted by spontaneously developing the quantum anomalous Hall state.

The formation and character of flat bands originating
from the $t_{2g}$ and $p$- orbitals highlight both the deep connection 
through orbital symmetry and a stark contrast in the material realization. 
In correlated oxide interfaces~\cite{Wang_2011}, flat bands emerge from the $t_{2g}$ manifold 
as a consequence of strong quantum confinement, lattice distortion, and orbital ordering 
at the two-dimensional interface. 
These bands, often intertwined with spin-orbit coupling and intrinsic disorder, 
lead to massively enhanced electron correlations and are implicated in emergent phenomena 
like interface magnetism and possible superconductivity~\cite{karmakar2025correlatedelectronsflatbands}.   
While $t_{2g}$-based flat bands offer a complex,
``real-world” platform where correlations compete
with other degrees of freedom, the $p$-orbital flat bands
provide a pristine, minimalist quantum simulator to validate
fundamental theories, both united by the shared
role of anisotropic orbital wavefunctions in generating
quenched dispersion and strong interaction effects.

\subsection{Itinerant ferromagnetism with $p$-orbitals}

A particularly intriguing application of these orbital optical lattices is the study of 
itinerant ferromagnetism in multi-orbital systems. In condensed matter physics, the emergence of ferromagnetism in itinerant electron systems has long been a subject of intense study, with the Stoner criterion providing the basic condition for ferromagnetic instability in the single-band systems. 
Nevertheless, the Stoner criterion neglects correlation effects and thus overestimates the ferromagnetism tendency. In fact, even under very strong repulsions, electrons in solids often remain paramagnetic.
In multi-orbital systems like those described by $t_{2g}$ orbitals that we have discussed in Sec.~\ref{sec3}, 
the situation becomes considerably more complex and rich. 
The orbital degrees of freedom introduce new channels for exchange interactions, 
particularly through the Hund's coupling, which favors parallel alignment of spins 
across different orbitals on the same atomic site.

In the $p$-orbital optical lattices, 
researchers can engineer multi-orbital systems 
where atoms occupy different $p$-orbital states ($p_x$, $p_y$, $p_z$) simultaneously. 
The anisotropic nature of orbital hopping, combined with carefully tuned interatomic interactions, 
creates conditions favorable for itinerant ferromagnetism through several distinct mechanisms. 
Isomorphic models as Eq.~\eqref{eq4} with the form of the multi-orbital extended Hubbard model 
can be constructed for these p-orbital ultracold atom systems on the relevant 
optical lattices. Exact results of itinerant ferromagnetism on square and cubic lattices 
were established in the limit of infinite $U$, and the ferromagnetism is expected to extend
to the strong coupling regime~\cite{PhysRevLett.112.217201}. 
This phenomenon bears striking resemblance to the itinerant ferromagnetism observed in $t_{2g}$ systems 
such as SrRuO$_3$~\cite{RevModPhys.84.253}, where the complex interplay between orbital and spin degrees of freedom stabilizes 
the ferromagnetic state. 
The optical lattice platform allows researchers to systematically 
map out the phase diagram of this orbital-driven ferromagnetism 
as a function of interaction strength, filling factor, and lattice geometry, 
providing crucial insights into the underlying mechanisms.

\subsection{Orbital homology in the Mott Regime}

The profound equivalence between the electronic structures of $p$- and $t_{2g}$-orbital manifolds 
extends beyond single-particle band topology into the deeply correlated Mott insulating regime, 
as exemplified by the emergence of the quantum compass model in both systems. 
For the $p$-band Mott insulators, it can be 
demonstrated that the spinless fermions confined  
to the anisotropic $p_x$ and $p_y$ orbitals of a two-dimensional optical lattice give rise, 
in the strong-coupling limit, to a highly directional superexchange interaction~\cite{PhysRevLett.100.200406}. 
This interaction, derived from the stark anisotropy of the orbital-lobe-directed hopping integrals, maps  
onto a quantum compass model. 
For a bond along a general direction of $\hat{e}_\phi = \cos \phi \hat{e}_x + \sin \phi \hat{e}_y$, 
the exchange has a compass form as
\begin{equation}
H_{ex} =  J_{\parallel} [ \vec{\tau} (\vec{r}) \cdot \hat{e}_{2\phi} ][ \vec{\tau} (\vec{r} + \hat{e}_{\phi} ) \cdot  \hat{e}_{2\phi}],
\label{orbitalexchange}
\end{equation}
where, the pseudospin vector operator ${\vec{\tau} = (\tau^x, \tau^y) }$ 
represents the two-fold orbital degrees of freedom, and the exchange anisotropy 
is directly frozen-in by the orbital geometry, leading to frustrated 
interactions and complex orbital ordering on different lattice geometries.

When applying the orbital exchange Hamiltonian Eq.~\eqref{orbitalexchange} 
to the honeycomb lattice, it leads to a quantum compass model.
It has a close connection to the Kitaev interaction in the Kitaev 
model where the exchange of each bond
is essentially Ising-like and bond-dependent. 
It is noted that, the Kitaev interaction
can be realized with the $j_{\text{eff}} =1/2 $ local moment 
of the $t_{2g}$ electrons subject 
to strong spin-orbit coupling (see Sec.~\ref{sec2})~\cite{Caobook,PhysRevLett.102.017205}.  
Here in Eq.~\eqref{orbitalexchange}, the Ising axes can be just chosen 
as the bond orientation by using a set of suitable basis. 
The compass model is not exactly solvable, but is heavily frustrated. 
Its classical ground states are mapped into configurations of the fully-packed 
loop model with an extra $U(1)$ rotation degree of freedom. 
Quantum fluctuations select a six-site plaquette ground state ordering 
pattern in the semiclassical limit from the ``{\it order from disorder}'' mechanism.

If the residing $p$-orbital fermion carries the spin, 
the resulting model should involve the spin exchange as well,
and the model is often known as Kugel-Khomskii spin-orbital exchange model
in the condensed matter community. 
 The very same effective models were known to govern the low-energy physics of certain $t_{2g}$-based 
 or $e_g$-based
 transition metal oxides in their Mott insulating state 
 for a long history~\cite{maekawa2004physics,KK1982,PTPS.160.155}.
This direct correspondence underscores that the $p$-$t_{2g}$ homology is not merely a single-particle coincidence but a robust principle that persists into the core of strongly correlated physics. 
It provides a foundational justification for using the pristine, 
tunable $p$-orbital optical lattices to quantum simulate the complex orbital 
ordering and magnetic frustration phenomena observed in correlated $t_{2g}$-electron materials.

The connection to $t_{2g}$ physics becomes particularly evident 
when considering the role of orbital frustration in promoting magnetic ordering. 
As we have remarked, in both $p$-orbital optical lattices and $t_{2g}$ electron systems, 
the directional nature of orbital hopping can lead to 
strong frustration~\cite{maekawa2004physics,PhysRevLett.100.200406}, 
where competing interactions prevent the system from finding a unique ground state. 
This frustration, combined with Hund's coupling that favors parallel spin alignment across orbitals, 
can drive the system toward a ferromagnetic state as one way to relieve the orbital frustration. 
Optical lattice experiments could explore this physics by engineering lattice geometries that 
enhance orbital frustration and studying the resulting magnetic correlations through quantum gas 
microscopy and other advanced detection techniques.

Furthermore, the optical lattice platform enables the investigation of how spin-orbit coupling influences 
itinerant ferromagnetism in multi-orbital systems. 
By introducing artificial gauge fields or laser-induced tunneling methods, 
researchers can engineer synthetic spin-orbit coupling that mimics the relativistic effects in $t_{2g}$ materials. 
This capability is particularly valuable for understanding materials like Sr$_2$RuO$_4$~\cite{PhysRevLett.101.026406}, 
where spin-orbit coupling plays a crucial role in determining the magnetic and superconducting properties. 
The combination of orbital degrees of freedom, tunable interactions, 
and synthetic spin-orbit coupling in optical lattices creates a powerful simulator 
for exploring the rich phase diagrams of $t_{2g}$ systems, 
including the competition between magnetic, superconducting, and orbital-ordered phases.

\subsection{Orbital quantum simulator}

The insights gained from $p$-orbital optical lattice experiments 
have significant implications for our understanding of the $t_{2g}$ materials. 
The study of orbital-driven ferromagnetism in optical lattices 
offers new perspectives on the long-standing problem of itinerant ferromagnetism in multi-orbital systems, 
suggesting that orbital degrees of freedom may play a more crucial role than previously appreciated.
Other orbital-related physics, such as
the orbital-selective Mott transitions~\cite{PhysRevLett.92.216402}, 
that has been proposed to
explain the peculiar properties of certain transition metal compounds with $t_{2g}$ orbitals,
may be clearly demonstrated 
with the $p$-orbital optical lattice systems.
The $p$-orbital optical lattices serve as useful quantum simulators for $t_{2g}$ physics, 
offering unparalleled control and measurement capabilities. 
The study of multi-orbital phenomena in these systems 
not only advances our fundamental understanding of orbital physics but also provides 
valuable guidance for the exploration and design of novel quantum materials 
with tailored properties. The functional equivalence
and orbital homology between $p$ and $t_{2g}$ orbital systems, 
so clearly demonstrated in these artificial quantum simulators, 
continues to reveal deep connections between seemingly disparate 
physical platforms and promises to 
illuminate the path toward new discoveries in correlated electron physics.

\section{Conclusion: Toward a Unified Orbital Framework for Quantum Materials}
\label{sec5}


The recognition of the orbital homology between the $p$ and $t_{2g}$ orbital systems   
represents more than just an interesting theoretical correspondence. It heralds a 
useful guiding scheme in how we conceptualize, design, and discover quantum materials. 
This orbital homology provides one unifying language that transcends traditional boundaries 
between different classes of materials and opens new vistas 
for controlling quantum matter. This framework offers a useful recipe for 
deciphering the complex interplay between orbital character, topology, 
and correlation effects across the quantum materials landscape.

The implications of this unified perspective point 
toward a future where materials design follows first principles guided 
by the orbital symmetry and topology. 
One can imagine the quantum materials engineered with the atomic precision, 
where the $p$-$t_{2g}$ equivalence and homology serve as the foundational principle 
for creating the heterostructures with certain topological properties, 
superconducting states with designed symmetry breaking, 
and magnetic systems with chosen frustration. 
The ability to translate insights between $p$-orbital and $t_{2g}$ systems 
creates a powerful useful loop, discoveries in complex transition metal compounds 
can inform the design of simpler $p$-orbital devices, while clean measurements in the
$p$-orbital platforms can validate theoretical predictions for the $t_{2g}$ materials. 
Although $t_{2g}$ systems are usually more correlated than some $p$-orbital systems 
and the Mott physics is often inaccessible with the $p$ orbitals,
this cross-pollination would still promise to accelerate the development of the 
quantum materials' research despite the potential differences between these 
two orbital systems that can be interesting on their own.

This orbital framework suggests several transformative research directions. 
First, it calls for an ``orbital mapping" of quantum materials
that is a comprehensive 
classification of electronic systems based on their orbital characteristics rather 
than their chemical composition alone. 
Such a classification would reveal hidden connections between seemingly disparate materials 
and predict new functional relationships. 
This could possibly lead to a new ``quantum materials genome" project~\cite{Bhattacharya_2023,Pablo}, 
where materials are characterized and designed according to their orbital genetic 
code, the specific combination of orbital symmetries, spin-orbit coupling strengths, 
and correlation effects that determine their quantum properties.
Second, it points toward the development of ``orbital engineering'' as a distinct discipline, 
where materials are designed specifically to optimize desired orbital characteristics for 
particular applications, much like the bandgap engineering in the semiconductor technology. 
Third, it suggests new pathways for discovering unconventional superconductors, 
quantum spin liquids, and other exotic states by leveraging the guiding principles 
revealed through the orbital homology.

The experimental implications are equally profound. 
Advanced synthesis techniques, from molecular beam epitaxy of oxide heterostructures 
to optical lattice engineering with ultracold atoms, can now be guided by this unified orbital principle. 
The ability to create ``orbital heterostructures'', where materials with complementary 
orbital characteristics are combined to produce emergent phenomena, 
represents an exciting frontier. Similarly, the time-resolved spectroscopies could 
track the orbital dynamics across different material systems, 
revealing possibly universal behaviors in how orbitals 
respond to external perturbations and participate in phase transitions.

As we advance into this new perspective, the $p$-$t_{2g}$ homology 
serves as both a guiding light and a call to action. It challenges us to 
rethink material categories, to develop new theoretical tools that 
embrace orbital complexity, and to create experimental platforms 
that exploit the orbital degrees of freedom. 
Most importantly, it reminds us that beneath the staggering diversity of quantum materials 
lies an underlying unity, a set of universal principles governed by symmetry and topology  
that transcend specific material implementations. 
By embracing this unified orbital perspective, we open the door to a future 
where quantum materials are not just discovered, but truly designed by atoms
and by orbitals to meet the technological challenges.


Finally, it is crucial to recognize that the $p$-$t_{2g}$ homology, while powerful, 
does not imply that the two systems are physically identical in all circumstances. 
The equivalence strictly concerns the local transformation properties   
under the octahedral point group and the resulting effective 
${l=1}$ algebra for the $t_{2g}$ triplet. The spatial wavefunctions of
the $p$  
and $t_{2g}$ orbitals are actually different. The $p$ orbitals are quasi-1D,
with lobes pointing along the bond directions, whereas the
$t_{2g}$ orbitals have a more 2D character, 
with lobes lying in the planes perpendicular to the bonding axes. 
Consequently, the intersite hopping integrals, 
and hence the tight-binding band structures 
and exchange interactions depend sensitively on lattice geometry 
and the nature of chemical bonding. 
Therefore, while the $p$-$t_{2g}$ homology provides a valuable unifying framework 
and a powerful design principle, its application to real materials must always be 
accompanied by an analysis of the specific lattice structure, orbital geometry, 
and hopping topology. Recognizing both the unity and the distinctions is essential 
for a balanced understanding and for avoiding the unexpected oversimplifications.

 
\section*{Acknowledgments}

We thank G. Khaliullin and Mike Norman for suggestion. 
GC is supported by NSFC No.~92565110, No.~12574061,
and BJNSF with No.~F261004. 
CW is supported by the National Natural Science
Foundation of China under the Grant No. 12234016 and
the New Cornerstone Science Foundation. 

\bibliographystyle{apsrev}
\bibliography{reference}

\end{document}